\documentclass[fleqn,twoside]{article}
\usepackage{espcrc2}

\usepackage{graphicx}

\usepackage {hyperref}

\title{Spin fluctuations influence on quasiparticle spectrum of realistic p-d model}

\author{Maxim M. Korshunov\address[Kirensky]{L.V. Kirensky Institute of Physics Siberian Branch of Russian Academy of Science, Krasnoyarsk, 660036, Russia} \thanks{Corresponding author. Tel.: +7-3912-494556; fax: +7-3912-438923; e-mail: \href{mailto:mkor@iph.krasn.ru}{mkor@iph.krasn.ru}.}, Sergei G. Ovchinnikov \addressmark[Kirensky]\address{UNESCO Chair "New materials and technology", Krasnoyarsk State Technical University, 660074 Krasnoyarsk, Russia.}, Alexei V. Sherman \address[Tartu]{Institute of Physics, University of Tartu, Riia 142, 51014 Tartu, Estonia}}

\begin{document}

\begin{abstract}
In the present work the multiband p-d model for $CuO_2$-layer is treated.
It was shown that for the realistic set of parameters besides Zhang-Rice
two-particle singlet there is non-negligible contribution of two-particle
triplet state to the top of the valence band. Also shown, that to gain
quantitative agreement with experimental data the minimal approximation
should include the spin fluctuations beyond the Hubbard-I scheme.
Quasiparticle spectrum, obtained in this approximation, is in fairly good
agreement with ARPES data on Bi2212 High-$T_c$ compound.
\vspace{1pc}
\end{abstract}

\maketitle

Since the \textit{ab initio} band theories still have problems in calculations of
strongly correlated electron systems (SCES) properties, the model approach
is still valuable and preferable in this field. One of the most interesting
SCES phenomena is High-$T_c$ superconductivity (HTSC). To explain
superconductive phase of the cuprates we should start with a model that
properly describes normal paramagnetic phase. Good candidate is a 3-band p-d
model \cite{1,2} but this model omits $d_{z^2}$ - orbitals on Cu and $p_z$ -
orbitals on apical Oxygen; importance of these orbitals is shown
experimentally \cite{3,4}. This lack is absent in the multiband p-d model of
transition metal oxides \cite{5}. Calculations \cite{6,7} of the quasiparticle
dispersion and spectral intensities in the framework of this model with use
of Generalized Tight Binding Method (GTBM) \cite{8} are in very good agreement
with ARPES data on insulating compound $Sr_2CuO_2Cl_2$ \cite{9}.

Other fascinating feature of multiband p-d model is that the difference
$\varepsilon _{T}-\varepsilon _{S}$ between energy of two-particle
Zhang-Rice-type singlet $À_{1g}$ and two-particle triplet $^{3}B_{1g}$
depends strongly on various model parameters, particularly on distance of
apical oxygen from planar oxygen, energy of apical oxygen, difference
between energy of $d_{z^2}$-orbitals and $d_{x^2-y^2}$-orbitals \cite{10}.
For the realistic values of model parameters $\varepsilon
_{T}-\varepsilon _{S}$ is less or equal to 0.5 eV \cite{6,7} contrary to
the 3-band model with this value being about 2 eV. Latter case was
considered in \cite{11,12} but due to the large singlet-triplet splitting the
contribution of singlet-triplet excitations in low-energy physics was
negligible. The former case will be considered in this paper.

To take into account triplet states we have derived an effective Hamiltonian
for $CuO_2$-layer \cite{13}. Hamiltonian of this effective singlet-triplet
model has the form of the generalized $t-t'-J$ model, but has several
important features: i) the account of a triplet leads to renormalization of
exchange integral $J$, ii) the model is asymmetric for n- and p-type systems
(for n-type systems the usual $t-J$ model takes place while for p-type
superconductors with complicated structure on the top of the valence band
the singlet-triplet excitations plays an important role; the asymmetry of p-
and n-type systems is known experimentally \cite{14}), iii) evolution of the
system with doping is described only by changes in band structure while all
parameters are fitted in undoped case and therefore fixed for all doping
levels \cite{6,7}.

\begin{figure}[h]
\begin{center}
\includegraphics[width=0.5\textwidth]{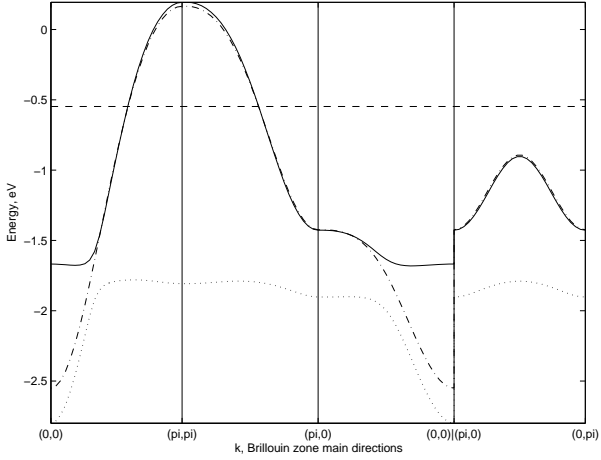}
\end{center}
\caption{Comparison of dispersion curves on top of the valence
band for effective singlet-triplet model (straight and dotted
lines) and $t-t'-J$ model (dash-dotted line), dashed line denotes
self-consistently obtained chemical potential.} \label{fig1}
\end{figure}

The paramagnetic non-superconductive phase was investigated in
Hubbard-I approximation both in the singlet-triplet and $t-t'-J$
models. Results for optimal doping (concentration of holes
$x=0.15$) are presented in fig.\ref{fig1}.

As one can see the mixture of triplet state (dotted line) and singlet state
(dashed line) is significant around (0,0) and ($\pi$,0) points. Meanwhile
singlet sub-band is rather wide, near 2 eV, that contradicts to experimental
ARPES data \cite{15}. It is a consequence of neglecting spin fluctuations in
Hubbard-I approximation \cite{16,17}. Because spin part of the effective
singlet-triplet model is the same as in the usual $t-J$ model, we can use in
our calculations the spin correlators self-consistently obtained in the
framework of $t-J$ model. By exploiting this approach, we were able to
calculate dispersion in paramagnetic non-superconductive phase of effective
singlet-triplet model for square lattice beyond the Hubbard-I approximation
similar to $t-J$ model \cite{16}. Spin correlations in $t-J$ model was calculated in
Rotationally Invariant Approximation \cite{17}, which gives close agreement
between calculated and experimental data on temperature and concentration
dependencies of magnetic susceptibility \cite{18}.

\begin{figure}[h]
\begin{center}
\includegraphics[width=0.5\textwidth]{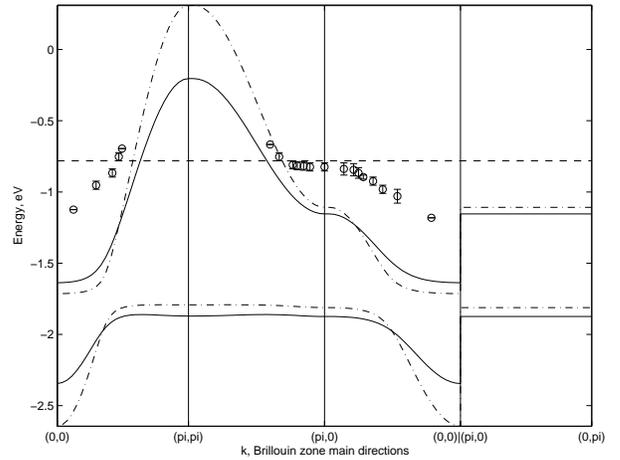}
\end{center}
\caption{Quasiparticle dispersion on top of the valence band for
effective singlet-triplet model in Hubbard-I approximation
(dash-dotted line), beyond the Hubbard-I approximation (straight
lines), chemical potential position (dashed line) and
experimental ARPES data (circles with error bars) \cite{15}.}
\label{fig2}
\end{figure}

Obtained dispersion for $x=0.15$ in this approximation together
with ARPES data on High-T$_{c}$ optimally doped compound Bi2212
\cite{15} is presented in fig.\ref{fig2}. The line-shape of
dispersion now is in good agreement with experimental data while
the positions of chemical potentials (and, consequently, Fermi
surfaces) quite differ. To reconcile our calculations and the
Fermi surface we have to shift the whole band by about 0.5 eV (at
optimal doping the position of flat region and chemical potential
should be nearly the same). There are few reasons why this shift
appears. One of the most important is the neglecting of 3-centers
terms during formulation of effective singlet-triplet model. While
these terms are very important for superconductive phase
\cite{19}, their influence on chemical potential renormalization
in normal phase may be significant too. Another necessary but yet
not used in this approach ingredient is the second and third
nearest neighbors. The importance of these hopping terms was
pointed out in work \cite{20}, where the LDA bands for 8-band,
3-band and 1-band Hamiltonians were compared.

This work has been supported by INTAS grant 01-0654, RFBR grant 03-02-16124,
Russian Federal Program "Integratsia" grant B0017, Program of Physical
Branch of RAS ``Strongly correlated electrons'' and ETF grant No. 5548.

\end{document}